\documentclass[twocolumn]{revtex4b5}
\usepackage{graphicx}
\usepackage{amsmath}
\begin{document}
\hsize\textwidth\columnwidth\hsize\csname@twocolumnfalse
\endcsname

\title{Charge sensing in carbon nanotube quantum dots on microsecond timescales}
\author{M. J. Biercuk$^1$, D. J. Reilly$^{1,2}$, T. M. Buehler$^2$, V. C.
Chan$^2$, J. M. Chow$^1$, R. G. Clark$^2$, C. M. Marcus$^1$}
\affiliation{$^1$ Department of Physics, Harvard University Cambridge, MA 02138}
\affiliation{$^2$ Centre for Quantum Computer Technology, School of Physics,
University of New South Wales, Sydney 2052, Australia}

\begin{abstract} We report fast, simultaneous charge sensing and transport
measurements of gate-defined carbon nanotube quantum dots. Aluminum radio
frequency single electron transistors (rf-SETs) capacitively coupled to the
nanotube dot provide single-electron charge sensing 
on microsecond timescales. Simultaneously, rf reflectometry allows fast
measurement of transport through the nanotube dot. 
Charge stability diagrams for the nanotube dot in the Coulomb blockade regime
show extended Coulomb diamonds into the high-bias regime, as well as even-odd
filling effects, revealed in charge sensing data.
\end{abstract}

\maketitle

% Intro
Carbon nanotubes are promising systems on which to base coherent electronic
 devices \cite{mceuen_prl99, liang_nature01, biercuk_prl05}. Due to a
combination of strong confinement, quantized phonon spectrum
\cite{hone_science00}, and zero nuclear spin, carbon nanotubes are likely to
exhibit long-lived coherent states. Key to the success of this technology is the
ability to manipulate electron states within a nanotube and perform fast,
efficient readout.
 Recent advances in device fabrication \cite{biercuk_prl05, javey_nature02,
javey_nature03, biercuk_nano04, biercuk_nano05} allow
 the creation of multiple quantum dots along the length of a tube with
controllable coupling by applying voltage biases to electrostatic top-gates.
However, readout of these structures has been limited to dc transport, which is
invasive and slow compared to 
relevant coherence times  \cite{forro_00}.

%Figure 1
\begin{figure}[t!]
\begin{center}
\includegraphics[width=5.5cm]{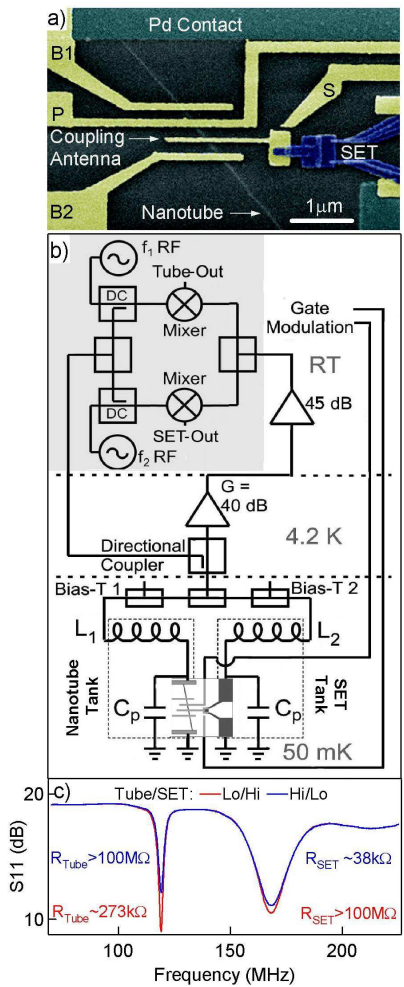}
\caption{{\bf a)} False-color SEM image of a representative nanotube quantum dot
device with integrated rf-SET. The nanotube is visible under $Al_2O_3$ and the
Pd contact (top).  Gates (yellow) are labeled on the figure.  The rf-SET (blue)
is aligned to a coupling antenna running over the nanotube. {\bf b)} Schematic
of the measurement setup for the multiplexed rf-reflectometry. Shaded area is
the demodulation circuit. {\bf c)} Reflected rf signal as measured with a
network analyzer for different values of nanotube  and SET resistances. In this
trace, the nanotube resistance is controlled with the back-gate while that of 
the SET is changed by shifting the bias voltage in or out of the superconducting
gap.}
\vspace{-0.5cm}
\end{center}
\end{figure}

In this Letter, we describe the integration of superconducting aluminum radio
frequency single electron transistors (rf-SETs) \cite{schoelkopf_science98} with
carbon nanotube quantum dot devices defined by electrostatic gates
\cite{biercuk_nano05}.  The rf-SET serves as a
 sensitive electrometer \cite{schoelkopf_nature00} and, when capacitively
coupled to the nanotube dot, provides a means of non-invasively detecting its
charge state on short timescales and in regimes not accessible with transport
measurements \cite{dicarlo_prl04, lu_nature03, buehler_apl05, vandersypen_prl04,
 elzerman_nature04}.  These represent the first charge sensing
 experiments with nanotube quantum dots using integrated charge
 detection. 
In addition, we make use of radio-frequency (rf) reflectometry that enables fast
transport
 measurements of the nanotube correlated with fast charge sensing by the
capacitively coupled
 rf-SET.  Previous work \cite{li_nano04} has used reflectometry to investigate
the microwave impedance of nanotubes 
 and transistor action at low (Hz) frequencies. The present study advances
previous work by simultaneously carrying out charge sensing together with rf
transport as well as by gaining five orders of magnitude in time resolution.
 
Carbon nanotubes were grown from patterned Fe catalyst islands on a Si/SiO$_2$
wafer via chemical
 vapor deposition.  Single-walled tubes with diameters less than $\sim 4$nm were
identified using atomic force microscopy and selectively contacted via electron
beam lithography
\cite{biercuk_prl05}.
  Contacts made from $\sim 15$ nm of Pd \cite{javey_nature03} were connected to
larger metallic
 pads defined by optical lithography. The entire device was then coated with
$\sim$ 35nm Al$_2$O$_3$ using 
 low-temperature atomic layer deposition (ALD). Three top-gates were then
aligned to each nanotube:
 two ``barrier gates'' (B1, B2) to deplete the underlying nanotube, defining a
quantum dot, with a third ``plunger gate'' (P) between them to tune the energy
of the dot (Fig. 1a)
\cite{biercuk_nano05}. The doped Si wafer also serves as a global back-gate. 
Nanotubes that showed little gate response (presumably metallic) were not
studied further.  The SET  island and the nanotube dot were capacitively coupled
by a 50nm Ti/AuPd (20\AA /30\AA) antenna which crosses
 the tube and sits under the SET island. The aluminum SET was fabricated using
double-angle evaporation on top of the coupling antenna (Fig 1a). 
Devices were mounted on a circuit board with rf coplanar waveguides and cooled
in a dilution refrigerator with a base temperature of 30-50\, mK. Electron
temperature measured in a similar configuration was in the range 100-200\,mK. 
Data from two integrated rf-SET/nanotube devices showing similar behavior are
reported.

Charge sensing is performed by monitoring the resistance of the SET using
rf-reflectometry. In the same way, direct transport measurements of the nanotube
are made using a small ($\mu V$) ac signal at rf frequencies (above 100MHz).
A schematic of the setup, showing the generation of the
reflectometry `carrier' signals at frequencies $f_1$ and $f_2$, is shown in
Fig.~1b. 
The two carrier signals are combined onto a single transmission line and fed to a
directional
 coupler at the 1K state of the dilution refrigerator. Two tank circuits
transform the high
 resistance of the SET ($\sim$ 50k$\Omega$) or nanotube ($\sim$ 200k$\Omega$)
towards $\sim50\Omega$, at the
 resonance frequencies $f_{1,2}$ set by the parasitic capacitance $C_p$ and
series
 chip-inductor ($L=780$nH for the nanotube and $L=330$nH for the SET). At
resonance, changes
 in resistance of either the nanotube or SET modify the $Q$-factor of the
respective tank circuit and the amount of reflected
 rf-power. After amplification at 4K (40dB) and room temperature (45dB) the
signals are homodyne detected using two mixers
and two local oscillators. 
Low-pass filtered output voltages from each mixer are proportional to the
change in respective device resistance. The use of frequency-domain
multiplexing allows both the SET and nanotube to be monitored using a common
transmission line and cryogenic amplifier
\cite{stevenson_apl02, buehler_jap04}. Bias-tees on the circuit board enable
standard dc transport measurements of both devices.

Figure 1c shows reflected power $S11$ from the tank circuits as a function of
frequency measured with a network analyzer after amplification. The two
resonances are identified at $f_1 \sim$ 120MHz for the nanotube and $f_2 \sim$
165MHz for
 the SET. Bandwidths are $ \sim 1$ MHz for the
 nanotube and $\sim 10$ MHz for the SET.

Figure 2a shows a charge stability plot for the SET
used for all measurements in Figs.~2 and 3. Plotted is the demodulated voltage
as a function of both the dc source-drain bias $V_{SD}^{SET}$ across the SET
and the voltage applied to a nearby gate. The SET is typically operated at the
threshold for
 quasi-particle transport, $V_{SD}^{SET} \sim 4\Delta/e$ ($\Delta$ is the
superconducting gap), where the rf-SET sensitivity is maximized.  Similar rf-SET devices measured in this setup exhibited charge sensitivities better than $\delta
q=10^{-5}e/\sqrt{Hz}$ \cite{buehler_jap04}.

We form a quantum dot in the carbon nanotube by applying appropriate voltages
 to gates B1 and B2 (Fig. 1a) with the back-gate set such that the nanotube is
n-type.
 The section of the nanotube between depletion regions formed by gates B1 and
 B2 serves as the quantum dot.  In this configuration, Coulomb blockade (CB) is
observed
 using standard low frequency lock-in measurements, as a series of conductance
peaks 
 as a function of P-gate voltage \cite{biercuk_nano05} (Fig. 2b).  
 
The energy of the dot is changed on fast timescales by 
 applying a triangle-wave voltage ramp to the P-gate. A compensating gate ramp
is applied to the S-gate to maintain the SET at a fixed
 conductance value. When the P- and S-gates sweep together in the same
direction, the SET is uncompensated and exhibits Coulomb blockade. In the region
where the P- and S-gates sweep in opposite directions
 the SET is compensated and may be held at a position of maximum
transconductance.

In the compensated configuration, the SET senses a characteristic sawtooth
charging pattern associated with P-gate induced 
tunneling of electrons onto the dot. The period of the sawtooth in P-gate
voltage is consistent with that measured directly from low frequency lock-in
measurements of CB in the nanotube.
 By contrast, if the barrier gate voltages are set such that there is no dot
formed
 in the tube, we observe a smooth line in the SET response (black trace, Fig.
2c).
 This indicates that the observed sawtooth response corresponds to charging of
 the gate-defined nanotube quantum dot. The magnitude of the charge signal
induced on the SET with the addition of a
 single electron to the nanotube dot is $\sim 0.2e$, indicating strong coupling
between the nanotube and the SET electrometer.

%Figure 2
\begin{figure}[t!]
\begin{center}
\includegraphics[width=7cm]{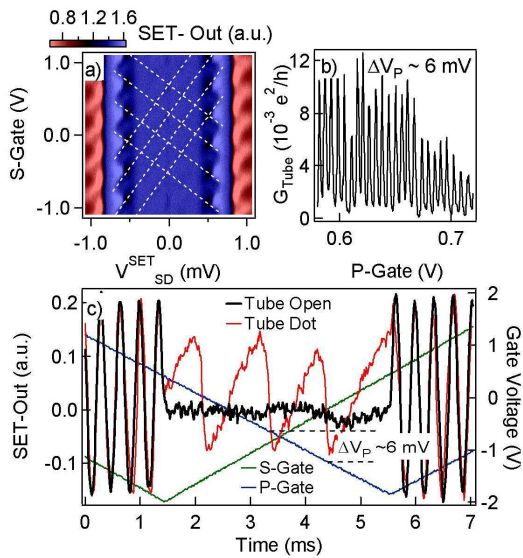}
\caption{{\bf a)} Charge stability diamonds for a superconducting rf-SET.
 Plotted on the intensity axis is the demodulated signal SET-Out. 
 Red corresponds to $\sim$ 50k$\Omega$ and blue $\sim$ 100M$\Omega$. Dotted
lines indicate conditions for resonant Cooper-pair tunneling.  {\bf b)} Coulomb
blockade in the nanotube as measured using standard lock-in techniques with
gates B1 and B2 near 2V, back gate = 18V, $V_{SD}$=1.5mV. $\Delta V_{P}$ is the
CB peak spacing in gate-voltage. {\bf c)} SET-Out
 signal (averaged 60 times, left axis) in the time domain (arb. offset) with the
SET biased to a sensitive region.  Green and blue traces are the applied
triangle wave gate ramps for S- and P-gates (before -40dB of attenuation)
 respectively (right axis). When no dot is formed in the nanotube (B1=B2=10V) we
observe a flat line in the compensated rf-SET
 response, while forming a dot as in panel b) yields a sawtooth charge sensing
signal.}
\vspace{-0.5cm}
\end{center}
\end{figure}

%Figure 3
\begin{figure}[t!]
\begin{center}
\includegraphics[width=7.6cm]{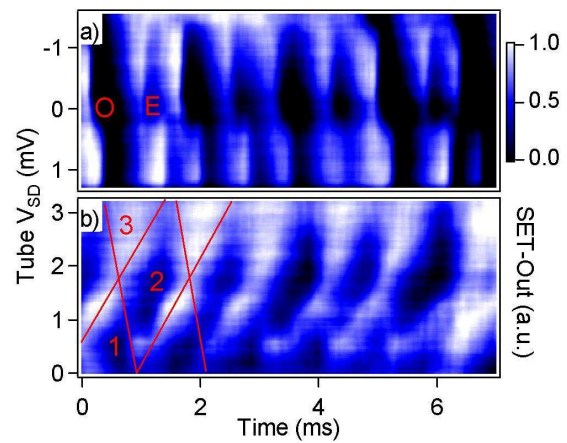}
\caption{{\bf a)} SET signal as a function of time (with a P-gate ramp applied)
and $V_{SD}$ across the nanotube showing even (E) and odd (O) filling of energy
states in the nanotube quantum dot. {\bf b)} Similar measurements in a different
configuration of B1 and B2 showing first, second and third order Coulomb
diamonds with increasing $V_{SD}$.  Red lines are guides to the eye indicating
the boundaries of diamonds in which charge number on the dot is fixed.}
\vspace{-0.5cm}
\end{center}
\end{figure}

Plotting the (compensated) SET-Out signal as a function of time (as the P-gate
voltage is ramped)
 and source-drain voltage $V_{SD}$ across the tube reveals the familiar diamond
pattern associated with  Coulomb
 blockade (Fig. 3a).  Here the applied voltage $V_{SD}$ across the nanotube also
couples capacitively to the SET itself. This effect is nulled by adding a
compensating dc offset to the gate ramp. 
 The nanotube dot charge configuration is fixed in the diamond regions (and
current blocked), while the blockade is lifted and current flow allowed at
 sufficiently high values of $V_{SD}$.  In appropriate biasing configurations of
gates B1
 and B2, we observe even-odd filling in the nanotube quantum dot
\cite{cobden_prl98}, indicated by
 an alternating pattern of large and small diamonds.  This is consistent with a
shell-filling model in which a single electron can enter a discrete energy level
 in the dot with charging energy $E_c=e^2/2C$ and quantum level spacing $\Delta
E$ ($C$ is the total dot capacitance).
 A second electron, with opposite spin to the first can enter the same orbital
state
 requiring only $E_c$.  Estimating $\Delta E$ for a dot of the size used in this
experiment
 to be  $\Delta E =750 \mu V$ using $E=hv_F/2L$, where $L\sim 1\mu m$ is the dot
length,
 is consistent with experimental measurements.  We have also observed four-fold 
 shell filling \cite{liang_prl02, sapmaz_prb05} in these gated nanotube devices.

%Figure 4
\begin{figure}[t!]
\begin{center}
\includegraphics[width=7.6cm]{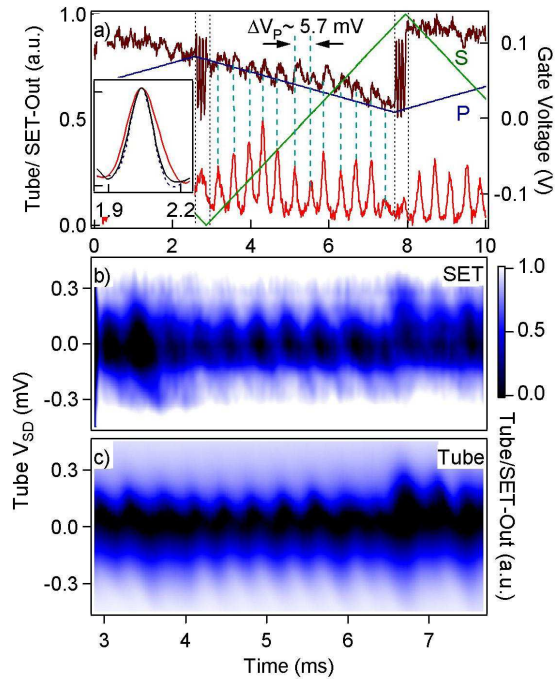}
\caption{{\bf a)} Fast, simultaneous measurement of the rf-SET and nanotube
using rf-reflectometry
 at tube $V_{SD} \sim 250 \mu$V. CB peaks are evident in the nanotube signal
(red lower trace) corresponding to a sawtooth in SET-Out. CB is also evident in
the SET signal at points where the gate biases change sweep direction and the
SET is uncompensated (within the dashed lines). 
B1, B2 $\sim$ 0V, back gate = 7.78V. {\bf Inset:} (lower left corner) shows the
backaction dependence of a CB peak in the tube as measured using reflectometry
at different $V_{SD}^{SET}$: biased to the gap (black trace), biased to the double Josephson quasiparticle peak (blue and dashed trace) and $\sim$ 3mV (red trace).   {\bf b)-c)}  A logarithmic
intensity plot of SET-Out and Tube-Out respectively (gate ramps identical to
those in panel a)) as a function of $V_{SD}$ across the tube. Coulomb diamonds
are visible in both panels, with key features reproduced between both. Each
sweep at fixed $V_{SD}$ has been averaged 1000 times.}
\vspace{-0.5cm}
\end{center}
\end{figure}

In addition to the low-bias diamonds commonly visible in transport, charge
sensing enables 
detection of the Coulomb staircase in $V_{SD}$ when one of the tunnel barriers
is made much larger than the other. In this configuration consecutive Coulomb
levels are populated from the source with increasing $V_{SD}$ before tunneling
to the drain can occur. Detection is possible because the SET senses the charge
state of the nanotube dot and not the current that
 flows from source to drain, which can be immeasurably small when the resistance
of one of the barriers is made large enough to observe the Coulomb staircase. In
the SET sensing signal we observe 
 diamonds centered at $V_{SD}=e/2C$, the bias
 corresponding to the apex of the  first order charging diamonds. First, second,
and the beginning of
 third order diamonds are visible in Fig. 3b, each offset by $e/2C$ from the
center
 of the diamonds of the next lower or higher order \cite{footnote1}.

By monitoring the demodulated signals from the tube and SET simultaneously,  we
can correlate rf-transport and charge sensing. 
Figure 4a shows both the demodulated signal from the
 nanotube together with SET-Out for S- and P-gate ramps with the nanotube
 in the CB regime (different device from Figs. 2 and 3).  CB peaks are evident
in the signal from the tube, and a sawtooth
 pattern is visible in SET-Out, with sequential charge addition occurring on
time scales
 of $\sim 300 \mu s$ (we have performed similar measurements with charge
addition
 periods $\sim 30 \mu s$, but systematic noise increased with gate speed). The
two signals are correlated as expected, with the apex of each CB peak falling
roughly in the middle of the charging sawtooth.  Further, the width of
 the sharp transition region for each sawtooth is roughly equivalent in time to
the width of the CB peak.

We have also studied how the $V_{SD}^{SET}$ biasing point of the SET influences
the
 Coulomb blockade in the nanotube quantum dot.  Consistent with measurements
made on Al  single-electron boxes \cite{turek_condmat05}, we observe asymmetries
and changes in the width of the CB peaks with varying $V_{SD}^{SET}$ across the
SET (Inset Fig. 4a). This behavior is likely due to
 a combination of heating  \cite{krupenin_prb05} and the backaction connected
with charge fluctuations of the SET island as current flows from source to
drain. We see a very slight {\it narrowing} of the
Coulomb blockade peaks in the nanotube
 dot when the rf-SET is biased near the double Josephson quasi-particle
peak (DJQP) \cite{clerk_prl02},
 relative to the CB peak-width when the SET is biased into the superconducting
gap. Separating back-action and heating effects will requires further study.

Charge stability plots for the
nanotube quantum dot, constructed from both SET-Out (Fig. 4b) and Tube-Out
(Fig. 4c) as a function of $V_{SD}$ across the nanotube, show 
 the nanotube (peaks) and SET (sawtooth)  signals to be correlated.  The rf-SET,
however, is sensitive to charge
 fluctuations in regions of $V_{SD}$ and P-gate voltage where direct transport
 measurements on the tube do not yield measurable currents, and where resistance
changes in the nanotube mapped through reflected-rf are immeasurable.

The authors wish to thank D. Barber, R. Starrett and N. Court for technical
assistance.  This work was supported by ARO/ARDA (DAAD19-02-1-0039 and -0191 and
DAAD19-01-1-0653),
 NSF-NIRT (EIA-0210736), and Harvard Center for Nanoscale Systems. M.J.B.
acknowledges support from an NSF
 graduate research fellowship and an ARO-QCGR fellowship.  D.J.R. acknowledges a
Hewlett-Packard postdoctoral fellowship.


\begin{thebibliography}{26}
\bibitem[{foo()}]{footnote1}
\bibinfo{note}{We also observe higher order diamonds without compensating for
the direct effect of $V_{SD}$ on the SET, but over a smaller range of gate
bias.}
\end{thebibliography}
\end{document}